\documentstyle[12pt]{article}
\newcommand{\beq}{\begin{equation}}
\newcommand{\eeq}{\end{equation}}
\newcommand{\bea}{\begin{eqnarray}}
\newcommand{\eea}{\end{eqnarray}}

\begin{document}
\setcounter{page}{0}
\topmargin 0pt
\oddsidemargin 5mm
\renewcommand{\thefootnote}{\fnsymbol{footnote}}
\newpage
\setcounter{page}{0}
\begin{titlepage}

\begin{flushright}
QMW-PH-95-26\\
{\bf hep-th/yymmnn}\\
June $14th$, $1995$
\end{flushright}
\vspace{0.5cm}
\begin{center}
{\Large {\bf Interacting Wess-Zumino-Novikov-Witten Models}} \\
\vspace{1.8cm}
\vspace{0.5cm}
{\large
Chris
Oleg A.
Soloviev
\footnote{e-mail: O.A.Soloviev@QMW.AC.UK}}\\
\vspace{0.5cm}
{\em Physics Department, Queen Mary and Westfield College, \\
Mile End Road, London E1 4NS, United Kingdom}\\
\vspace{0.5cm}
\renewcommand{\thefootnote}{\arabic{footnote}}
\setcounter{footnote}{0}
\begin{abstract}
{We study the system of two WZNW models coupled to each other via the
current-current interaction. The system is proven to possess the
strong/weak
coupling duality symmetry. The strong coupling phase of this theory
is
discussed in detail. It is shown that in this phase the interacting
WZNW models
approach non-trivial conformal points along the renormalization group
flow. The
relation between the principal chiral model and interacting WZNW
models is
investigated.}
\end{abstract}
\vspace{0.5cm}
\centerline{June 1995}
 \end{center}
\end{titlepage}
\newpage
\section{Introduction}

String theory symmetries act on the space of string solutions that
from the
point of view of the world-sheet description can be observed as
transformations
of the coupling constants of underlying two-dimensional quantum field
theories.
Therefore, some information about local properties of these
symmetries can be
extracted by infinitesimal variations of the given string solution.
In other
words, one can make use of perturbation theory methods in order to
study the
string symmetries. It may seem that the so-called duality symmetries
of string
theory will fail to be detected within this perturbative approach.
However, in
the vicinity of the self-dual points one can still carry on with a
proper
perturbative expansion \cite{Gates}. Of course, the problem now will
be how to
identify the self-dual points? Apart from the last issue,
perturbation theory
appears to be an adequate tool in elaborating string symmetries.

Some of the perturbations of a given two-dimensional conformal field
theory
corresponding to a certain string solution may take us away from the
space of
string solutions. This will happen when the perturbation breaks the
conformal
invariance. Therefore, in order to reveal some of the string
symmetries, it may
be necessary to extend the space of string solutions. Then,
restricted to
string solutions, these symmetries either will be seen as discrete or
will not
be seen at all.

The hope is that by integrating infinitesimal transformations it will
be
possible to reconstruct the entire group of string symmetries.
Correspondingly,
this may turn out to bring us to a better understanding of string
field theory
itself.
Luckily, one may find a two-dimensional model which corresponds to a
general
element of the entire string symmetry group or its subgroup. In the
present
paper
we shall argue that such a general element can be described as a
system of two
WZNW models coupled to each other through a current-current
Thirring-like
interaction. As yet we have not proven this conjecture, but we have
found some
supporting evidence for it.

The importance of interacting WZNW models also arises as one tries to
find a
unified description of two-dimensional integrable models. The latter
have
proven to be extremely useful in modeling many physical situations
and
phenomena. At present it
seems that these models may be of great importance in understanding
non-trivial properties of string theory, for example, S-duality
\cite{Schwarz}. So far only the principal chiral model has been
considered in
this context \cite{Schwarz}. In spite of its merits, this model, in
many cases,
appears to be an approximation, which does not always allow one to
grasp some
of the essential effects of a system it is supposed to describe. For
example,
the hidden affine symmetry of the non-linear sigma model \cite{Dolan}
is not
sufficient for describing the conformal Sugawara construction, whose
proper
description is achieved in terms of the Wess-Zumino-Novikov-Witten
(WZNW) model
\cite{Knizhnik}. One can observe the latter as a modified PCM
\cite{Witten}.
However, it seems to be more appropriate to regard the
PCM as a modified WZNW theory \cite{Soloviev1}.

The way in which the PCM stems from the WZNW model suggests that
there may exist a larger class of two-dimensional field theories
which possess
the important property of integrability. We shall discuss this
relation
between the PCM and the WZNW model in section 3. It turns
out that a broad class of two-dimensional integrable theories can be
described
in terms of interacting WZNW models. The PCM corresponds to the
isoscalar current-current interaction \cite{Soloviev1}. More general
current-current interactions give rise to more general integrable
models. Our
hope is that the space of the given integrable models may form a
connected
multitude parametrized by the Thirring coupling constants. This space
of
theories might turn out to be of help in understanding the space of
string
compactifications and solutions, which ought to be presented by CFT's
in this
space. The geometry and topology of this model space may play a role
in
formulating the background independent string field action.

This paper is organized as follows. In section 2, two interacting
level $k_1$
and $k_2$ WZNW models are introduced and their classical properties
are
studied. In section 3, the PCM  is presented as a system of two
interacting
WZNW models. In section 4, we exhibit the strong/weak coupling
duality of
interacting WZNW models at the quantum level. In section 5, by using
this
duality, we analyze the strong coupling phase of the theory in
question. We
show that the theory has a non-trivial critical point in this phase.
In section
6, we discuss the non-perturbative conformal points of the system of
two
interacting WZNW models. We conclude in section 7.

\section{Classical properties of interacting WZNW models}

We would like to start with some classical properties of interacting
WZNW
models, which then will be of help in developing quantum theory.

Let $S_{WZNW}(g_1,k_1)$ and $S_{WZNW}(g_2,k_2)$ be the actions of two
WZNW
models of level $k_1$ and $k_2$, and with $g_1$ and $g_2$ taking
values in the
Lie groups $G_1$ and $G_2$ respectively:
\begin{eqnarray}
S_{WZNW}(g_1,k_1)&=&{-k_1\over4\pi}\left\{\int\mbox{Tr}|g^{-1}_1\mbox{
d}g_1|^2~+~
{i\over3}\int\mbox{d}^{-1}\mbox{Tr}(g^{-1}_1\mbox{d}g_1)^3\right\},\nonumber\\
&
& \\
S_{WZNW}(g_2,k_2)&=&{-k_2\over4\pi}\left\{\int\mbox{Tr}|g^{-1}_2\mbox{
d}g_2|^2~+~
{i\over3}\int\mbox{d}^{-1}\mbox{Tr}(g^{-1}_2\mbox{d}g_2)^3\right\}.\nonumber
\end{eqnarray}
These models are invariant under the following affine symmetries
\begin{eqnarray}
g_1&\to&\bar\Omega_1(\bar z)g_1\Omega_1(z),\nonumber\\ & & \\
g_2&\to&\bar\Omega_2(\bar z)g_2\Omega_2(z),\nonumber\end{eqnarray}
where the parameters $\Omega_{1,2},~\bar\Omega_{1,2}$ are arbitrary
independent
functions of $z$ and $\bar z$ respectively.

It turns out that there is an interaction between these two WZNW
models which
preserves the affine symmetries of the free theories. This
interaction is given
as follows
\begin{equation}
S_I={-k_1k_2\over\pi}\int d^2z\mbox{Tr}^2(g^{-1}_1\partial g_1~S~\bar
\partial g_2g^{-1}_2),\end{equation}
with the coupling $S$ belonging to the direct product of two Lie
algebras
${\cal G}_1\otimes{\cal G}_2$. The symbol $\mbox{Tr}^2$ indicates a
double
tracing over the indices of a matrix from the tensor product
${\cal G}_1\otimes{\cal G}_2$. We shall assume that $S$ is
invertible, so that
$\dim G_1=\dim G_2$.

We call the system described by the following action
\begin{equation}
S(g_1,g_2,S)=S_{WZNW}(g_1,k_1)~+~S_{WZNW}(g_2,k_2)~+~S_I(g_1,g_2,S)
\end{equation}
a system of two interacting WZNW models.
Remarkably, when $S\ne0$, the interacting system still has $\hat
G^L_1\otimes
\hat G^R_1\otimes\hat G^L_2\otimes\hat G^R_2$ affine symmetry
\cite{Soloviev},
under which
\begin{eqnarray}
g_1&\to&\bar\Omega_1(\bar z)g_1h_1\Omega_1(z)h^{-1}_1,\nonumber\\ & &
\\
g_2&\to&h_2^{-1}\bar\Omega_2(\bar
z)h_2g_2\Omega_2(z),\nonumber\end{eqnarray}
where $h_1,~h_2$ are non-local functions of $g_1,~g_2$ satisfying
\begin{eqnarray}
\bar\partial h_1h^{-1}_1&=&2k_2\mbox{Tr}S\bar\partial
g_2g_2^{-1},\nonumber\\ &
&\\
h_2^{-1}\partial h_2&=&2k_1\mbox{Tr}Sg^{-1}_1\partial
g_1.\nonumber\end{eqnarray}
The $\bar\Omega_1$ and $\Omega_2$ transformations remain local, while
the
$\bar\Omega_2$ and $\Omega_1$ transformations are now intrinsically
non-local,
as they involve $h_1,~h_2$. The local $\bar\Omega_1$ and $\Omega_2$
symmetries
are manifest, whereas the proof of the remaining non-local ones is
given in
\cite{Hull}.

The fact that the interacting theory possesses the affine symmetry
amounts to
the existence of an infinite number of affine currents and, as a
consequence of
it, to the integrability of the theory. Like in the ordinary WZNW
model, the
affine currents of interacting WZNW models emerge via the equations
of motion.
Indeed, the equations of motion of the theory in eq. (2.4) can be
written as
\begin{equation}
\partial\bar{\cal J}_{(1)}=0,~~~~~~~~~\bar\partial{\cal
J}_{(2)}=0,\end{equation}
where
\begin{eqnarray}
\bar{\cal J}^{\bar a}_{(1)}&=&\bar J^{\bar a}_{(1)}~+~2k_2\phi^{a\bar
a}_1S^{a\bar b}\bar J^{\bar b}_{(2)},\nonumber\\ & & \\
{\cal J}^a_{(2)}&=&J^{a}_{(2)}~+~2k_1\phi^{a\bar
a}_2S^{b\bar a}J^{b}_{(1)},\nonumber\end{eqnarray}
with
\begin{eqnarray}
J^a_{(1)}&=&-{k_1\over2}\mbox{Tr}(g^{-1}_1\partial
g_1~t_1^a),\nonumber\\
J^a_{(2)}&=&-{k_2\over2}\mbox{Tr}(g^{-1}_2\partial
g_2~t^a_2),\nonumber\\
\bar J^{\bar a}_{(1)}&=&-{k_1\over2}\mbox{Tr}(\bar
\partial g_1 g_1^{-1}~t^{\bar a}_1),\nonumber\\ & & \\
\bar J^{\bar a}_{(2)}&=&-{k_2\over2}\mbox{Tr}(\bar
\partial g_2 g_2^{-1}~t^{\bar a}_2),\nonumber\\
\phi_1^{a\bar a}&=&\mbox{Tr}(g_1t^a_1g_1^{-1}t^{\bar
a}_1),\nonumber\\
\phi_2^{a\bar a}&=&\mbox{Tr}(g_2t^a_2g_2^{-1}t^{\bar
a}_2),\nonumber\end{eqnarray}
where $t^i_{1,2}$ are the generators of the Lie algebras ${\cal
G}_{1,2}$
associated with the Lie groups $G_{1,2}$,
\begin{eqnarray}
\left[t^i_1,t^j_1\right]&=&f^{ij}_{(1)k}t^k_1,\nonumber\\ & & \\
\left[t^i_2,t^j_2\right]&=&f^{ij}_{(2)k}t^k_2,\nonumber\end{eqnarray}
with $f^{ij}_{(1,2)k}$ the structure constants.

The analytical currents $\bar{\cal J}_{(1)},~{\cal J}_{(2)}$
correspond to the
two local $\bar\Omega_1$ and $\Omega_2$ symmetries respectively.
There are in
addition two non-local conserved currents ${\cal J}_{(1)},~\bar{\cal
J}_{(2)}$
corresponding to the two non-local $\bar\Omega_2$ and $\Omega_1$
symmetries.
These two local and two non-local conserved currents generate an
infinite
number of conserved charges that results in the integrability of the
system of
two interacting WZNW models with ${\it arbitrary}$ coupling matrix
$S$. For the
local currents $\bar{\cal J}_{(1)},~{\cal J}_{(2)}$ an infinite set
of
conserved charges can be built up according to
\begin{eqnarray}
\bar{\cal J}_{(1)n}&=&\oint{d\bar z\over2\pi i}~\bar z^n\bar{\cal
J}_{(1)}(\bar
z),\nonumber\\ & & \\
{\cal J}_{(2)n}&=&\oint{d z\over2\pi i}~ z^n{\cal J}_{(2)}(
z).\nonumber\end{eqnarray}
With respect to the classical Poisson bracket the charges
$\bar{\cal J}_{(1)n},~{\cal J}_{(2)n}$ form the two affine algebras
$\hat{\bar{\cal G}}_1$ and $\hat{\cal G}_2$ at levels $k_1$ and
$k_2$,
respectively,
\begin{eqnarray}
\left[\bar{\cal J}^{\bar a}_{(1)n},\bar{\cal J}^{\bar
b}_{(1)m}\right]&=&f^{\bar a\bar b}_{(1)\bar c}~\bar{\cal J}^{\bar
c}_{(1)n+m}~+~{k_1\over2}\delta^{\bar a\bar
b}m\delta_{n+m,0},\nonumber\\ & &
\\
\left[{\cal J}^{ a}_{(2)n},{\cal J}^{
b}_{(2)m}\right]&=&f^{a b}_{(2) c}~{\cal J}^{
c}_{(2)n+m}~+~{k_2\over2}\delta^{a
b}m\delta_{n+m,0}.\nonumber\end{eqnarray}

The point to be made is that the existence of the analytical currents
$\bar{\cal J}_{(1)},~{\cal J}_{(2)}$ follows directly from the
equations of
motion. Therefore, in quantum theory, these currents will exist as
long
as the equations of motion can be promoted to the quantum level.
Hence,
whenever the system of interacting WZNW models exists as a quantum
theory, it
will be automatically integrable.

There is another interesting classical property of interacting WZNW
models.
Namely, given a system of two interacting WZNW models one can find a
dual
theory \cite{Soloviev1}. The procedure is as follows. First, we
introduce
auxiliary Lie-algebra-valued variables $Q,~\bar P$ and write the
interaction
action in the (first-order) form
\begin{eqnarray}
S(g_1,g_2,Q,\bar
P)&=&S_{WZNW}(g_1,k_1)~+~S_{WZNW}(g_2,k_2)\nonumber\\ & & \\
&-&{1\over2\pi}\int d^2z\mbox{Tr}(k_1g^{-1}_1\partial g_1\bar
P~+~k_2\bar
Q\bar\partial g_2g_2^{-1}~-~{1\over2}Q\cdot S^{-1}\cdot\bar
P).\nonumber\end{eqnarray}
The requirement of invertibility of $S$ comes from the last equation.
If one
eliminates $Q,~\bar P$ via their equations of motion, one regains
(2.4).
Instead, we express $Q,~\bar P$ in terms of new variables $h_1,~h_2$
as
\begin{equation}
\bar\partial h_1h_1^{-1}=\bar P,~~~~~~~~~h^{-1}_2\partial
h_2=Q\end{equation}
and use the Polyakov-Wiegmann formula \cite{Wiegmann} to obtain
\begin{eqnarray}
S(g_1,g_2,Q,\bar P)&=&S_{WZNW}(g_1h_1,k_1)~+~S_{WZNW}(h_2g_2,k_2)
{}~+~S_{WZNW}(h_1,-k_1)
\nonumber\\ &
& \\
&+&S_{WZNW}(h_2,-k_2)~+~{1\over4\pi}\int
d^2z\mbox{Tr}^2h^{-1}_2\partial h_2 S^{-1}\bar\partial
h_1h^{-1}_1.\nonumber\end{eqnarray}
If we introduce new variables
\begin{equation}
\tilde g_1=g_1h_1,~~~~~~~~\tilde g_2=h_2g_2,\end{equation}
then we can see that $\tilde g_1,~\tilde g_2$ decouple from
$h_1,~h_2$ and the
action for the latter is the interacting
system of two level $-k_1,~-k_2$ WZNW models with the coupling
\begin{equation}
\tilde S=-{1\over4k_1k_2}S^{-1}.\end{equation}
At the quantum level the given duality undergoes some quantum
modifications
which will be the subject of section 4.

We have exhibited that the system of interacting WZNW models is
integrable and
possesses the duality symmetry. In our further discussion we shall
study these
classical properties for quantum theory.

\section{The PCM from WZNW models}

For the further purposes it is instructive to demonstrate a curious
relation
between the PCM and the interacting system
of WZNW models we have described in the previous section.

Let us consider the system of two interacting identical WZNW models
with the
isoscalar coupling matrix
\begin{equation}
S=s~I,\end{equation}
where $I$ is the identity in ${\cal G}\otimes{\cal G}$, $s$ is some
parameter.
In this case, the affine currents in eqs. (2.8) take the form
\begin{eqnarray}
\bar{\cal J}_1=\bar J_1~+~2ksg_1\bar J_2g_1^{-1},\nonumber\\ & & \\
{\cal J}_2=J_2~+~2ksg_2^{-1}J_1g_2.\nonumber\end{eqnarray}
Correspondingly the equations of motion (2.7) can be rearranged as
follows
\begin{eqnarray}
\bar\partial J_1~-~{4s\over1+2ks}\left[J_1,\bar
J_2\right]&=&0,\nonumber\\ & &
\\
\partial\bar J_2~-~{4s\over1+2ks}\left[J_1,\bar
J_2\right]&=&0.\nonumber
\end{eqnarray}

It is convenient to introduce new variables
\begin{equation}
A=J_1,~~~~~~~~~\bar A=-\bar J_2\end{equation}
to rewrite eqs. (3.20) in the form
\begin{eqnarray}
\bar\partial A~+~\lambda~\left[A,\bar A\right]&=&0,\nonumber\\ & & \\
\partial\bar A~-~\lambda~\left[A,\bar
A\right]&=&0,\nonumber\end{eqnarray}
where
\begin{equation}
\lambda={4s\over1+2ks}.\end{equation}
The given equations are nothing but the equations of motion of the
PCM (see
e.g. \cite{Schwarz}). Thus, we have exhibited the classical
equivalence between the latter and the system of two interacting WZNW
models.
In particular, this equivalence allows us to apply the results in
Ref.\cite{Schwarz} to the interacting WZNW models. It might be
interesting
to connect the affine symmetries in eqs. (2.5) with the hidden affine
symmetries of the PCM \cite{Schwarz}.

The PCM was shown to correspond to the isoscalar
interaction between two identical WZNW theories. In general the
system of two
interacting WZNW models provides a uniform description to a large
class of
integrable theories some of which might turn out to be new.

\section{The quantum duality}

Because duality symmetries occur to play an important role in
physics, we
shall discuss in detail the duality symmetry of the system of two
interacting
WZNW models at the quantum level. Later on we shall make use of this
symmetry
to solve the theory in the strong coupling regime.

Let us consider the partition function of the system of two
interacting WZNW
models
\begin{equation}
Z(k_1,k_2,S)=\int{\cal D}g_1{\cal
D}g_2\exp[-S(g_1,g_2,S)].\end{equation}
It is necessary to point out that the definition of the function
integral in
eq. (4.24) in general depends on the coupling constant matrix $S$.
For example,
there is a critical point (Polyakov-Wiegmann conformal point) at
which the
system of interacting WZNW models acquires the gauge symmetry.
Therefore, the
functional integral at this point has to be further defined in the
Faddeev-Popov
manner.

As in the classical case, we introduce the auxiliary variables under
the
functional integral to obtain
\begin{eqnarray}
Z(k_1,k_2,S)&=&J^{-1}\int{\cal D}g_1{\cal D}g_2{\cal D}\bar P{\cal
D}Q\exp[-S(g_1,g_2, Q,\bar P)],\nonumber\\ & & \\
J&=&\int{\cal D}\bar P{\cal D}Q\exp\left[-{1\over4\pi}\int
d^2x\mbox{Tr}\bar
PS^{-1}Q\right].\nonumber\end{eqnarray}
Here the quantity $J$ depends only on topology of the world sheet.
Integration
over $Q,~\bar P$ in the partition function $Z$ amounts to using the
equations
of motion for $Q,~\bar P$ giving rise to the equivalence of the two
expressions
in (4.24) and (4.25). The step further from the variables $Q,~\bar P$
to the
variables $h_1,~h_2$ entails some new features compared to the
classical
theory. Namely, this change of variables is accompanied by the
Jacobians and
the local counterterms which affect the interaction between $h_1$ and
$h_2$.
All of these modifications can be properly taken into account via the
following
formula
\begin{eqnarray}
Z(k_1,k_2,S)&=&J^{-1}~Z^{(1)}_{WZNW}(k_1)~Z^{(2)}_{WZNW}(k_2)~
\bar Z^{(1)}_{ghost}
{}~Z^{(2)}_{ghost}\nonumber\\ & & \\
&\times&Z\left(-k_2-2c_V(G_2),-k_1-2c_V(G_1),-{A(k_1,k_2)S^{-1}\over
4(k_1+2c_V(G_1))(k_2+2c_V(G_2))}\right),\nonumber\end{eqnarray}
where $Z_{WZNW}^{(1,2)}(k_{1,2})$ are the partition functions of the
WZNW
models
on $G_{1,2}$ with levels $k_{1,2}$, whereas
\begin{eqnarray}
\bar Z^{(1)}_{ghost}&=&\int{\cal D}b{\cal D}c\exp\left[-\int
d^2z\mbox{Tr}_1b\bar\partial c\right],\nonumber\\ & & \\
Z^{(2)}_{ghost}&=&\int{\cal D}\bar b{\cal D}\bar c\exp\left[-\int
d^2z\mbox{Tr}_2\bar b\partial\bar c\right],\nonumber\end{eqnarray}
with $b,~c,~\bar b,~\bar c$ being Lie-algebra-valued ghost-like
fields arising
in the course of the change of variables in eqs. (2.14). The constant
$A(k_1,k_2)$ stands for the corrections due to the local
counterterms. We shall
fix $A$ from the consistency condition.

Note that the appearance of the ghosts is fairly natural because the
dual
theory involves WZNW models at negative level. Therefore, the
unitarity of the
whole system assumes making use of some BRST-like procedure for
ruling out
negative normed states. We shall discuss the unitarity of the model
under
consideration in some particular cases.

Let us turn to the consistency condition on the constant $A$. It
comes along
in the following way. By applying the duality transformation
repeatedly to the
dual theory, one can derive the following identity
\begin{eqnarray}
Z(k_1,k_2,S)&=&J^{-2}~Z^{(1)}_{WZNW}(k_1)~Z^{(2)}_{WZNW}(k_2)~
\bar Z^{(1)}_{ghost}
{}~Z^{(2)}_{ghost}\nonumber\\
&\times&Z^{(2)}_{WZNW}(k_2-2c_V(G_2))~Z^{(1)}_{WZNW}(k_1-2c_V(G_1))
{}~\bar Z^{(2)}_{ghost}~Z^{(1)}_{ghost}\\
&\times&Z\left(k_1,k_2,{A(-k_1-2c_V(G_1),-k_2-2c_V(G_2))(k_1+2c_V(G_1)
)
(k_2+2c_V(G_2))\over
A(k_1,k_2)k_1k_2}~S\right).\nonumber\end{eqnarray}
The crucial observation is that
\begin{eqnarray}
&Z^{(1)}_{WZNW}(k_1)&Z^{(2)}_{WZNW}(k_2)~Z^{(1)}_{WZNW}(k_1-2c_V(G_1))
{}~Z^{(2)}_{WZNW}(k_2-2c_V(G_2))\nonumber\\ & & \\
&\times&\bar Z^{(1)}_{ghost}~Z^{(1)}_{ghost}~\bar Z^{(2)}_{ghost}~
Z^{(2)}_{ghost}=Z_{G_1/G_1}~Z_{G_2/G_2},\nonumber\end{eqnarray}
where $Z_{G_{1,2}/G_{1,2}}$ are the partition functions of the
topological
conformal field theories \cite{Spiegelglas}. Thus, formula (4.28)
takes the
form
\begin{equation}
Z(k_1,k_2,S)=J^{-2}~Z_{G_1/G_1}~Z_{G_2/G_2}~Z(k_1,k_2,S'),\end{equation}
where
\begin{equation}
S'={(k_1+2c_V(G_1))(k_2+2c_V(G_2))A(-k_1-2c_V(G_1),-k_2-2c_V(G_2))\over
k_1k_2A(k_1,k_2)}~S.\end{equation}
We have shown in \cite{Hull} that $J$ can be expressed as follows
\begin{equation}
J^2=Z_{G_1/G_1}~Z_{G_2/G_2}.\end{equation}
Thus, the factor $J^{-2}$ cancels with the factor
$Z_{G_1/G_1}~Z_{G_2/G_2}$ and
we arrive at the following equality
\begin{equation}
Z(k_1,k_2,S)=Z(k_1,k_2,S').\end{equation}
The last formula implies that
\begin{equation}
S=S'\end{equation}
up to modular transformations in the space of coupling constants. All
in all we
find the consistency condition
\begin{equation}
{A(-k_1-2c_V(G_1),-k_2-2c_V(G_2))\over
A(k_1,k_2)}={k_1k_2\over(k_1+2c_V(G_1))(k_2+2c_V(G_2))}.\end{equation}

Apart from the modular symmetry, this consistency condition may have
many
solutions. To pick up the right one, one has to consider the
classical limit.
Taking into account eq. (2.15), it is not difficult to see that the
classical
limit amounts to the following condition
\begin{equation}
{\lim_{k_1,k_2\to\infty}}A(k_1,k_2)=1.\end{equation}
Bearing in mind this limit the constant $A$ is fixed without
ambiguities
\begin{equation}
A(k_1,k_2)=\left[{(k_1+2c_V(G_1))(k_2+2c_V(G_2))\over
k_1k_2}\right]^{1/2}.\end{equation}

Finally we write done the exact formula for the duality symmetry of
the system of two interacting WZNW models:
\begin{eqnarray}
Z(k_1,k_2,S)&=&J^{-1}~Z_{WZNW}^{(1)}(k_1)~Z_{WZNW}^{(2)}(k_2)~\bar
Z^{(1)}_{ghost}~Z_{ghost}^{(2)}\nonumber\\ & & \\
&\times&Z\left(-k_2-2c_V(G_2),-k_1-2c_V(G_1),-{S^{-1}\over4\sqrt{(k_1+
2c_V(G_1))
(k_2+2c_V(G_2))k_1k_2}}\right).\nonumber\end{eqnarray}

Nicely there is a check for eq. (4.38). Let us consider the case,
when
$k_1=k_2=k,~G_1=G_2=G$. The system of two identical interacting WZNW
models has
the Polyakov-Wiegmann conformal point at
\begin{equation}
S_{PW}={I\over2k},\end{equation}
where $I$ is the identity from ${\cal G}\otimes{\cal G}$. At this
point eq.
(3.38) reads off
\begin{eqnarray}
Z\left(k,k,{I\over2k}\right)&=&J^{-1}Z_{WZNW}^2(k)\bar
Z_{ghost}Z_{ghost}\nonumber\\ & & \\
&\times&Z\left(-k-2c_V(G),-k-2c_V(G),-{I\over2(k+2c_V(G))}\right).\nonumber
\end{eqnarray}
Now one can see that the PW conformal point of the original theory
goes to the
PW conformal point of the dual theory. This means that the both
theories
possess the gauge symmetry. Thus, we have to impose an appropriate
gauge
condition in
order to fix the given gauge arbitrariness. This can be done without
introducing
dynamical Faddeev-Popov ghosts by setting one of the two group
elements to the
identity matrix. After that one can evaluate the Virasoro central
charge of the
original CFT and the dual one. Based on it, one can verify that
equality (4.38)
persists for the Virasoro central charges.

Note that the dual theory in eq. (4.38) describes the interaction
between two
WZNW models with negative levels. Because of that, this theory does
not belong
to the space of interacting WZNW models with positive levels. Thus,
the duality
transformation links the two theories of different types. Remarkably,
the
strong coupling phase of one theory goes into the weak coupling phase
of its
dual one. In the next section we shall show how to make use of the
given
duality in exploring the strong coupling regime of the system of two
interacting WZNW models with positive levels.

\section{The strong coupling phase}

Somewhat to simplify the further consideration we put
$k_1=k_2=k,~G_1=G_2=G$.
The formulation of the system of two interacting WZNW models as given
in
section 2, turns out to be convenient in analysis of the weak
coupling phase.
Indeed, in this regime one can expand around the free WZNW models in
the
interaction, which is determined by a marginal conformal operator.
However,
when the coupling
matrix $S$ is large, one can no longer relay on perturbation in $S$.
Our aim
in this section is to reformulate the theory in eq. (2.4) so that it
will
be suitable for investigation of the strong coupling phase.

The idea is to make the following change of variables
\begin{equation}
g_1\to h(\tilde g_1)\cdot\tilde g_2,~~~~~~~~g_2\to\tilde
g_1,\end{equation}
where $\tilde g_1,~\tilde g_2$ are new variables, whereas the
function
$h(\tilde g_1)$ is the solution of the following equation
\begin{equation}
\partial hh^{-1}=-2k\mbox{Tr}S\tilde g_1^{-1}\partial\tilde
g_1.\end{equation}
This determines $h$ up to changes of the form
\begin{equation}
h\to h\Lambda(\bar z),\end{equation}
where $\Lambda$ is an antiholomorphic matrix function,
$\partial\Lambda=0$, and
we will pick some particular solution $h_0(z,\bar z)$. One can take
any other
solution $h_0\Lambda$. However, because the Jacobian of the change in
eq.
(5.41) is always equal to one, a new choice will not change the
theory.

By considering $\bar\partial(\partial hh^{-1})$ and using eq. (5.42)
we also
obtain
\begin{eqnarray}
\bar\partial h_0h_0^{-1}(z,\bar z)&=&-2k\mbox{Tr}S\tilde
g^{-1}_1\bar\partial_{\bar z}\tilde g_1(z,\bar z)\nonumber\\ & & \\
&+&2k\int d^2z\bar\partial_{\bar z}G(z,\bar z,y,\bar y)v(y,\bar
y),\nonumber\end{eqnarray}
where
\begin{eqnarray}
v(y,\bar y)&=&\mbox{Tr}S\tilde g_1^{-1}(y,\bar
y)\left[\bar\partial_{\bar
y}\tilde g_1(y,\bar y)\tilde g^{-1}_1,~\partial_y\tilde g_1(y,\bar
y)\tilde
g_1^{-1}(y,\bar y)\right]\tilde g_1(y,\bar y)\nonumber\\ & & \\
&+&\left[\mbox{Tr}S\tilde g^{-1}_1(y,\bar y)\partial_y\tilde
g_1(y,\bar
y),~\bar\partial h_0(y,\bar y)h^{-1}_0(y,\bar
y)\right]\nonumber\end{eqnarray}
and the Green function $G(z,\bar z;y,\bar y)$ satisfies
\begin{equation}
\bar\partial_{\bar z}\partial_z G(z,\bar z;y,\bar
y)=\delta(z,y)\delta(\bar
z,\bar y).\end{equation}
We regularize the Green function in such a way that
\begin{equation}
{\lim_{y\to z,\bar y\to\bar z}}\partial_zG(z,\bar z;y,\bar
y)=0\end{equation}
so that, despite its nonlocality, the right hand side of eq. (5.44)
is well
defined, even when $(y,\bar y)\to (z,\bar z)$.

Equations (5.42), (5.44) are sufficient to express $h_0$ in terms of
$\tilde
g_1$ and its derivatives. Note that the symmetry in eq. (5.43) gets
fixed when
one makes use of eq. (5.44).

Writing the action in terms of the new variables
$\tilde g_1,~\tilde g_2$ and using the Polyakov-Wiegmann formula, we
obtain
\begin{eqnarray}
S(g_1,g_2,k)&\to&S(\tilde g_1,\tilde g_2,k)=S_{WZNW}(\tilde g_2,k)~+~
S_{WZNW}(\tilde g_1,k)~+~S_{WZNW}(h_0,k)\nonumber\\ & & \\
&-&{k^2\over\pi}\int d^2z\mbox{Tr}\tilde g_1\partial\tilde
g_1~S~\bar\partial
h_0 h_0^{-1},\nonumber\end{eqnarray}
where $h_0$ is a nonlocal function of $\tilde g_1$ satisfying (5.32),
(5.34).
Remarkably, after this change of variables, the field $\tilde g_2$
completely
decouples from $\tilde g_1$.

The price we pay for the factorization is a highly nonlocal theory
for the
variable $\tilde g_1$. While $\tilde g_2$ is governed simply by a
WZNW action,
the action for $\tilde g_1$ is
\begin{eqnarray}
S(\tilde g_1)&=&S_{WZNW}(\tilde g_1,k)~+~S_{WZNW}(h_0(\tilde
g_1),k)\nonumber\\
& & \\
&-&{k^2\over\pi}\int d^2z\mbox{Tr}^2\tilde g^{-1}_1\partial\tilde
g_1~S~\bar\partial h_0(\tilde g_1)h^{-1}_0(\tilde
g_1),\nonumber\end{eqnarray}
which is non-local as $h_0$ is a non-local function of $\tilde g_1$.
In the
Wess-Zumino term in $S_{WZNW}(h_0(\tilde g_1),k)$, $h_0$ and $\tilde
g_1$ are
extended to functions of $z,~\bar z$ and an extra coordinate $t$ and
an
equation analogous to (5.44) can be found for
$\partial_th_0h_0^{-1}$. Then
$h_0$ appears in (5.49) only through its derivatives, so that (5.49)
can be
written in terms of $\tilde g_1$ using (5.42), (5.44) and the
analogous
equation for $\partial_t h_0h_0^{-1}$. The resulting action takes the
following
form
\begin{eqnarray}
S(\tilde g_1)&=&S_{WZNW}(\tilde g_1,k)~+~{k^3\over\pi}\int
d^2z\mbox{Tr}^2(\mbox{Tr}S\tilde g_1^{-1}\partial\tilde
g_1~\mbox{Tr}S\tilde
g_1^{-1}\bar\partial\tilde g_1)\nonumber\\ & & \\
&+&{k\over\pi}\int d^2z\bar\partial(\tilde g^{-1}_1\partial\tilde
g_1)\Psi~+~{\cal O}(S^3),\nonumber\end{eqnarray}
where
\begin{equation}
\Psi(z,\bar z)=k^2\int d^2yG(z,\bar z;y,\bar y)\mbox{Tr}S~v(y,\bar
y).\end{equation}

The important point to be made is that all interaction terms in eq.
(5.50) are
to be understood as normal ordered, so that all divergences coming
from
non-local terms do not occur.

For our purposes it is necessary to establish the duality symmetry of
the
non-local theory described by eq. (5.49). To this end, let us
consider the
following partition function
\begin{equation}
Z_B(k,S)=\int{\cal D}\tilde g_1\exp[-S(\tilde g_1)].\end{equation}
This emerges in the partition function of the system of interacting
WZNW models
\begin{equation}
Z(k,k,S)=Z_{WZNW}(k)Z_B(k,S).\end{equation}
Correspondingly formula (4.38) can be rewritten in terms of $Z_B$ as
follows
\begin{equation}
Z_B(k,S)=J^{-1}~Z_{WZNW}(k)~\bar
Z_{ghost}~Z_{ghost}~Z_{WZNW}(-k-2c_V(G))~Z_B(-k-2c_V(G),\tilde
S),\end{equation}
where $\tilde S$ is the dual coupling constant matrix
\begin{equation}
\tilde S={-S^{-1}\over4(k+2c_V)k}.\end{equation}
Taking into account observations (4.29) and (4.32) we find the
following
identity
\begin{equation}
Z_B(k,S)=Z_B(-k-2c_V,\tilde S).\end{equation}
This is the relation which manifests the precise equivalence between
the two
theories.

Apparently, when $S\to\infty,~\tilde S\to0$ and vice versa. Thus,
equality
(5.56) allows us to
link the strong-coupling regime of one theory to the weak-coupling
phase of
another. In other words, we have got the opportunity to explore the
quantum
field theory with the large coupling constant. Certainly, the dual
model is
very non-trivial. Nevertheless, it can be studied at least within
perturbation
theory \cite{Hull}.

It is worth reminding the method of studying the dual theory
\cite{Hull}. To
start with, we would like to point out some general aspects of this
theory in
the small $\tilde S$ limit. In this limit, the system of two
interacting WZNW
models can be regarded as a sum of two non-unitary WZNW models
perturbed by the
marginal (but not truly marginal) operator
\begin{equation}
X=J_2\otimes\bar J_1,\end{equation}
where
\begin{equation}
J_2={(k+2c_V)\over2}h^{-1}_2\partial h_2,~~~~~~~\bar
J_1={(k+2c_V)\over2}\bar\partial h_1h_1^{-1}.\end{equation}
Since $X$ is not a truly marginal operator, the beta function of
$\tilde S$
does not vanish for all $\tilde S$. At the same time, because $X$ is
marginal, the beta function starts with ${\cal O}(\tilde S^2)$
terms,
\begin{equation}
\beta={\cal O}(\tilde S^2)~+~...\end{equation}
Therefore, in order to check, for example, whether there are
non-trivial fixed
points or not, one has to go in computations to order $\tilde S^3$.
Technically
such computations seem to be quite involved as one has to expand to
$X^3$.
Fortunately, there is a trick which will allow us to overcome these
technical
obstacles. We shall show that the factorization simplifies the
calculation of
the beta function.

We proceed to make it clear that the limit $S\to\infty$ or $\tilde
S\to0$ does
exist, and we, indeed, can perform perturbation around $\tilde S=0$.
This is
important step in our discussion. According to identity (4.38) the
limit
$S\to\infty$ corresponds to the CFT with the Virasoro central charge
\begin{equation}
c_\infty=2\dim G.\end{equation}
This formula comes from the right hand side of eq. (4.38). On the
left hand
side of eq. (4.38), the limit $S\to\infty$ gives rise effectively to
the theory
\begin{equation}
S(S\to\infty)=-{k^2\over\pi}\int d^2z\mbox{Tr}^2(g^{-1}_1\partial
g_1~S~\bar\partial g_2g^{-1}_2).\end{equation}
Let us assume that $S=\rho~I$, where $\rho\to\infty$. Then, eq.
(5.61) takes
the form
\begin{equation}
S(\rho)=-{k^2\rho\over\pi}\int d^2z\mbox{Tr}(g^{-1}_1\partial
g_1\bar\partial
g_2g_2^{-1}).\end{equation}
The latter can be regarded as the non-linear sigma model
\begin{equation}
S(\lambda)=-{1\over4\pi\lambda^2}\int d^2z\mbox{Tr}(g^{-1}_1\partial
g_1\bar\partial g_2g_2^{-1})\end{equation}
taken in the limit $\lambda^2=1/(4k^2\rho)\to0$. In the given limit,
the
non-linear sigma model is equivalent to the CFT of the direct sum of
two
Abelian affine current algebras with dimensions equal to $\dim G$.
The Virasoro
central charge of the given CFT is equal to $2\dim G$. Thus, we
have shown that the limit $S\to\infty$ leads us to the same CFT as
the limit
$\tilde S\to0$. Therefore, this agreement justifies the existence of
the both
limits and their consistency to each other. This is a crucial
observation for
pursuing expansion around the CFT at $\tilde S=0$.

For the dual theory the factorized one is given by
\begin{eqnarray}
\tilde S(\tilde h_1)&=&S_{WZNW}(\tilde
h_1,-k-2c_V)~-~{(k+2c_V)^3\over\pi}\int
d^2z\mbox{Tr}^2(\mbox{Tr}\tilde S\tilde h^{-1}_1\partial\tilde
h_1~\mbox{Tr}\tilde S\tilde h^{-1}_1\bar\partial\tilde
h_1)\nonumber\\ & & \\
&-&{(k+2c_V)\over\pi}\int d^2z\bar\partial(\tilde
h^{-1}_1\partial\tilde
h_1)\tilde\Psi~+~{\cal O}(\tilde S^3).\nonumber\end{eqnarray}
The given non-local theory can be properly understood as follows. We
consider
the non-local term and all other higher in $\tilde S$
terms as being normal ordered with respect to the two first terms in
eq.
(5.64). Whereas the second term in eq. (5.64) is thought of as being
a
perturbation on the conformal WZNW model at level $-k-2c_V$. Such a
view of the
given non-local functional allows us to develop selfconsistent
expansion in the
small coupling $\tilde S$.

The equation of motion of the WZNW model is as follows
\begin{equation}
\bar\partial(\tilde h^{-1}_1\partial\tilde h_1)=0.\end{equation}
When the second term in eq. (5.64) is turned on, the equation of
motion gets
perturbed according to
\begin{equation}
\bar\partial(\tilde h^{-1}_1\partial\tilde h_1)={\cal O}(\tilde
S),\end{equation}
where the right hand side of eq. (5.66) is expressed in terms of
normal ordered
products of conformal operators of the CFT. Eq. (5.66) can be used
further for
normal ordering the non-local term and higher order terms in  eq.
(5.64). Now
it becomes clear that since the non-local term contains
$\bar\partial(\tilde
h^{-1}_1\partial\tilde h_1)$, its normal ordered expression will be
of order
$\tilde S^3$ due to the equation of motion (5.66). Therefore, if we
confine
ourselves to leading in $\tilde S$ orders, we can stick to the
following
approximation
\begin{equation}
\tilde S(\tilde h_1)=S_{WZNW}(\tilde
h_1,-k-2c_V)-{(k+2c_V)^3\over\pi}\int
d^2z\mbox{Tr}(\mbox{Tr}\tilde S\tilde h^{-1}_1\partial\tilde h_1~
\mbox{Tr}\tilde S\tilde h^{-1}_1\bar\partial\tilde h_1)+{\cal
O}(\tilde
S^3).\end{equation}
Remarkably, the given approximation is a local theory.

Now we want to exhibit that in the space of models described by eq.
(5.67)
there is a class of renormalizable theories with the coupling
constant matrix
$\tilde S$ having the following structure
\begin{equation}
\tilde S=\sigma~\hat S,\end{equation}
with $\sigma$ being a small parameter and $\hat S$ a fixed matrix
which is not
subject to renormalization. Eq. (5.68) can be understood as a line in
the space
of coupling constants.

With the coupling as in eq. (5.68) the theory in eq. (5.67) reads off
\begin{equation}
S(\tilde h_1)=S_{WZNW}(\tilde h_1,-k-2c_V)~-~\epsilon\int
d^2z~O^{\hat
S}(z,\bar z)~+~{\cal O}(\sigma^3),\end{equation}
where we have introduced the following notations
\begin{equation}
\epsilon={4(k+2c_V)\over\pi}\sigma^2,~~~~~~~~~~O^{\hat S}=\hat
S^{a\bar a}
\hat S^{b\bar a}:\tilde J^a\bar{\tilde J^{\bar b}}\tilde\phi^{b\bar
b}:
.\end{equation}
Here
\begin{eqnarray}
\tilde J&=&{(k+2c_V)\over2}\tilde h^{-1}_1\partial\tilde
h_1,\nonumber\\
\bar{\tilde J}&=&{(k+2c_V)\over2}\bar\partial\tilde h_1\tilde
h^{-1}_1,\\
\tilde\phi^{a\bar a}&=&\mbox{Tr}(\tilde h^{-1}_1t^a\tilde h_1t^{\bar
a}).\nonumber\end{eqnarray}

The operator $O^{\hat S}$ in eq. (5.69) is a Virasoro primary
operator with the
conformal weight
\begin{equation}
\Delta_O=1-{c_V\over k+c_V}.\end{equation}
It is transparent that $\Delta_O$ lies in the interval between 0 and
1 and,
thus, it is a relevant operator in the non-unitary WZNW model.
Despite the
non-unitarity of the WZNW theory, $O^{\hat S}$ belongs to the unitary
range of
the Kac-Kazhdan determinant and, hence, it provides a unitary
representation of
the Virasoro algebra.

In order for the theory in eq. (5.69) to be renormalizable, the
operator
$O^{\hat S}$ has to form a closed OPE algebra. The last condition
results in an
algebraic equation for the matrix $\hat S$ \cite{Soloviev2}.
Solutions to this
equation yield renormalizable relevant perturbations on the WZNW
model.

We shall show that the limit $\tilde S\to0$ is consistent with the
limit
$k\to\infty$. Therefore, for our goal it will be sufficient to
consider the
large $k$ solutions for $\hat S$. In this limit, there is one
invertible
solution given by \cite{Soloviev2}
\begin{equation}
\hat S^{a\bar a}\hat S^{b\bar a}={\delta^{ab}\over
c_V}~+~...,\end{equation}
where dots stand for the higher in $1/k$ corrections. With the given
$\hat S$
the theory is renormalizable and the beta function can be computed:
\begin{equation}
\beta(\epsilon)=(2-2\Delta_O)\epsilon~-~\pi~\epsilon^2~+~...\end{equation}
Note that as $\epsilon$ is proportional to $\sigma^2$, eq. (5.74)
agrees with
eq. (5.59).

Now it becomes clear that the beta function in eq. (5.74) has a
non-trivial
conformal point
\begin{equation}
\epsilon^{*}={4\over\pi k}~+~...\end{equation}
Recalling the relation between $\epsilon$ and $\tilde S$ given by eq.
(5.70)
one can see that the large $k$ limit is consistent with the limit
$\tilde
S\to0$. Furthermore, because of this relation, solution (5.75) gives
rise to
two
solutions for $\tilde S$. However, only one of these two critical
points can be
reached along the renormalization group flow \cite{Hull}. Indeed,
going to the
point $\tilde S^{*}>0$ necessitates the flow to
pass the Polyakov-Wiegmann critical point,
at which the system acquires the gauge symmetry. Therefore,
perturbation in the
direction to $\tilde S^{*}>0$ cannot be continuous. At the same time,
it might
be likely the case that these two
critical points are related by some duality symmetry \cite{Gates}.
The CFT's
corresponding to these two conformal points must be essentially
equivalent and,
hence, we can carry on with one of these points. From now on we shall
study the
fixed point $\tilde S^{*}<0$.

Known $\epsilon^{*}$ we find the critical point $\tilde S^{*}<0$ and,
correspondingly, a conformal point for the system of two interacting
WZNW
models
\begin{equation}
S^{*}={1\over2\sqrt2k}~+~...\end{equation}
This point does not coincide with the PW conformal point and is new.

Although, in the limit $k\to\infty$, the critical point $S^{*}\to0$,
it cannot
be seen by expanding the beta function around $S=0$. In fact, the
given
conformal point becomes visible only in the strong-coupling phase via
expansion
around $S=\infty$. The strong-coupling region is completely covered
by the
weak-coupling phase of the dual theory. Together, the weak-coupling
phase of
the system of two interacting WZNW models and the weak-coupling phase
of the
dual theory, cover the whole range of positive values of the coupling
constant
$s$ in eq. (3.18).




Note that at $S=\infty$ there is another fixed point with the
Virasoro central
charge given by eq. (5.60).

We can say more about the point $S^{*}$ in eq. (5.76) noticing that
the
perturbative CFT (5.69) at this point can be identified with an exact
CFT
\cite{Soloviev3}. The latter is nothing but the WZNW model with level
$k$,
\begin{equation}
S(\tilde h_1,S=S^{*})=S_{WZNW}(\tilde h_1,k).\end{equation}
By using this result, one can derive from eq. (5.54) the following
expression
\begin{equation}
Z(k,k,S^{*})=Z^2_{WZNW}(k).\end{equation}
This result is rather unexpected. It exhibits that in the space of
interacting
WZNW models there exists a point at which the interacting theory
becomes
entirely equivalent to the system of non-interacting WZNW models. In
other
words, the theory with the interaction suddenly becomes free again at
the
particular value of the coupling. The similar effect has been
described for the
system of spinor fields with the four-fermionic interaction
\cite{Dashen}. In
fact, the fixed point $S^{*}$ is related to the so-called isoscalar
Dashen-Frishman conformal point of the non-Abelian Thirring model
\cite{Hull}.

Due to the equivalence between the system of two interacting WZNW
models and
the PCM described in section 3, the weak-coupling phase of the later
is related
to the strong-coupling regime of the former. However, at the critical
point we
just have found above, the given equivalence has to break down, like
at the
point $S=0$ or at the Polyakov-Wiegmann conformal point. Exploring of
the PCM
in the vicinity of the Dashen-Frishman fixed point might be an
interesting
task.

In this section we analyzed the conformal property of the system of
two
interacting WZNW
models in the strong-coupling phase. Also we pointed out that
renormalizability of this theory admits many other $\hat S$'s to
exist, that
will lead to new critical points. In the next section we shall
discuss the
algebraic structure of the theory at hand to reveal the possibility
of
existence of many more non-perturbative conformal points.

\section{The current-current algebraic approach}

Based on the results obtained in the previous sections we want to
propose a
non-perturbative way of finding conformal points of the system of two
interacting WZNW models.

We have exhibited in the previous section that one critical point can
be found
in the large $k$ limit due to the duality. However, there may exist
some other
fixed points at particular values of $k$. Certainly if such critical
points
actually exist, they cannot
be found in the large $k$ limit. In order to discover these points,
one should
solve
the theory exactly that is, of course, an extremely difficult task.
However, we
shall show that the large $k$ limit
provides insight into the algebraic structure of the theory, which
turns out to
be a key point in the search for non-perturbative conformal points.

According to definition, at a conformal point the system is
characterized by
the
stress-tensor with the two analytic components $T$ and $\bar T$,
\begin{equation}
\bar\partial T=0,~~~~~~~\partial\bar T=0.\end{equation}
The third component which is equal to the trace of the stress-tensor
vanishes due to the conformal symmetry.
In addition, $T$ and $\bar T$ form two copies of the Virasoro
algebra.
Thus, whenever the theory possesses the conformal symmetry, it must
have two
analytic operators forming two Virasoro algebras.

At the classical level, the system of two interacting WZNW models is
conformal.
So is the non-local theory described by eq. (5.49) as well as its
dual one. Let
us consider the dual theory. The classical stress-tensor has the
following two
components
\begin{eqnarray}
T(\tilde S)&=&{(-1)\over k+2c_V}\tilde J^2~+~{(-1)\over k+2c_V}\tilde
J^2_{\tilde h_0}~+~4\tilde J(\mbox{Tr}\tilde S\tilde J_{\tilde
h_0}),\nonumber\\
& & \\
\bar T(\tilde S)&=&{(-1)\over k+2c_V}\bar{\tilde J^2}~+~{(-1)\over
k+2c_V}\bar{
\tilde
J^2}_{\tilde h_0}~+~4\bar{\tilde J}(\mbox{Tr}\tilde S\bar{\tilde
J}_{\tilde
h_0}),\nonumber\end{eqnarray}
where
\begin{eqnarray}
\tilde J&=&{(k+2c_V)\over2}\tilde h^{-1}_1\partial\tilde
h_1,\nonumber\\
\bar{\tilde J}&=&{(k+2c_V)\over2}\bar\partial\tilde h_1\tilde
h^{-1}_1,\\
\bar\partial\tilde h_0\tilde h^{-1}_0&=&2(k+2c_V)\mbox{Tr}\tilde
S\tilde
h^{-1}_1\bar\partial\tilde h_1.\nonumber\end{eqnarray}
Using the equation for $\tilde h_0$, the holomorphic component $T$
can be
altered to the local form in terms of $\tilde h_1$:
\begin{equation}
T(\tilde S)=\left({-\delta^{ab}\over k+2c_V}~+~4(k+2c_V)\tilde
S^{a\bar
a}\tilde S^{b\bar a}\right)\tilde J^a\tilde J^b.\end{equation}
This is curious, because
\begin{equation}
\bar\partial\tilde J\ne0,\end{equation}
whereas
\begin{equation}
\bar\partial T(\tilde S)=0.\end{equation}
In other words, the non-analytic part of the function $\tilde
J(z,\bar z)$ does
not contribute into the holomorphic component of the classical
stress-tensor.
Therefore, if we take $\tilde J(z,\bar z)$ at the origin, i.e. at
$z=\bar z=0$,
then there will be no evolution in the $\bar z$-direction for the
quantity
$T(\tilde S)$.

At $z=\bar z=0$, we have
\begin{equation}
T(0)=\left({-\delta^{ab}\over k+2c_V}~+~4(k+2c_V)\tilde S^{a\bar
a}\tilde
S^{b\bar a}\right)\tilde J^a(0,0)\tilde J^b(0,0).\end{equation}
At all other moments of time, we must have
\begin{equation}
T(z)=\left({-\delta^{ab}\over k+2c_V}~+~4(k+2c_V)\tilde S^{a\bar
a}\tilde
S^{b\bar a}\right)\hat J^a(z)\hat J^b(z),\end{equation}
where
\begin{equation}
\hat J^a(z)\equiv\tilde J^a(z,0).\end{equation}

Let us assume that $\tilde
J^a(z,\bar z)$ form the current algebra with respect to the Poisson
brackets
\begin{equation}
\{\tilde J^a(z,\bar z),~\tilde J^b(y,\bar z)\}=f^{ab}_c\tilde
J^c(z,\bar
z)\delta(z,y)~+~central~term.\end{equation}
At the classical level $\tilde J$ transforms as a primary field under
the
conformal group. Correspondingly, the algebra in eq. (6.88) is
invariant under
evolution in the $\bar z$-direction. Therefore, it will hold for
$\hat J(z)$ as
well,
\begin{equation}
\{\hat J^a(z),~\hat J^b(y)\}=f^{ab}_c\hat
J^c(z)\delta(z,y)~+~central~term.\end{equation}
Unfortunately, algebra (6.88) is inconsistent with the construction
in eq.
(6.82) for arbitrary $\tilde S$. It has been proven in \cite{Halpern}
that the
current algebra can be consistent with the Virasoro algebra only at
some
special values of the matrix $L_{ab}$ in the construction
\begin{equation}
T=L_{ab}~\hat J^a\hat J^b.\end{equation}
Whereas in the case under consideration, $T(\tilde S)$ forms the
Virasoro
algebra with arbitrary $\tilde S$. Hence, at the classical level, the
quantity
$\hat J$ has to have a more complicated algebra which has to be
consistent with
the classical Virasoro algebra for arbitrary $\tilde S$. In this
context, it
might be interesting to know, whether the algebra found in
\cite{Bardakci} is
the algebra forming by $\hat J$.

The situation may change in quantum theory. Indeed, the classical
stress-tensor, in the course of quantization, is subject to
renormalization
even at conformal points \cite{Dashen},\cite{Knizhnik}. Therefore,
the
quantum
consistency condition between the current algebra and the Virasoro
algebra can
be different from the classical consistency condition due to
renormalization.

In order to see how the renormalization works, we turn to the
conformal point
in the large $k$ limit. It has been shown in the previous section
that the
dual non-local theory in eq. (5.64) flows to the WZNW model with
level $k$,
\begin{equation}
\tilde S(\tilde S^{*})=S_{WZNW}(k).\end{equation}
In other words, the stress-tensor in eqs. (6.80) takes the
affine-Sugawara form
at the given conformal point,
\begin{equation}
\bar T(\tilde S^{*})={1\over k+c_V}:\bar{\tilde{\cal
J}^2}:,\end{equation}
where $\bar{\tilde{\cal J}}$ is the affine current of the WZNW model
with level
$k$. The last expression exhibits that the classical stress-tensor
$T(\tilde
S)$ gets renormalized to the Sugawara stress-tensor at the point
$\tilde
S^{*}$. The current $\bar{\tilde{\cal J}}$ coincides with the
renormalized
antiholomorphic current
\begin{equation}
\bar{\tilde{\cal J}}=\bar{\tilde J}~-~2(k+2c_V)\tilde
h_1(\mbox{Tr}\tilde
S\bar{\tilde J_{\tilde h_0}})\tilde h^{-1}_1.\end{equation}

At the classical level this current forms the affine algebra with
level
$-k-2c_V$. In the course of quantization $\bar{\tilde{\cal J}}$
continues to be
analytic due to the equation of motion. However, it undergoes some
renormalization which affects the central charge. It turns out that
in the
large $k$ limit it is not hard to compute the renormalized central
charge of
the renormalized affine current. By using perturbation theory we find
\begin{equation}
k_R=-k~+~2k^3\sigma^2~+~...\end{equation}
At the critical point $S^{*}$, the renormalized central charge $k_R$
is given
by
\begin{equation}
k_R(S^{*})=-k~+~2k~+~...=k,\end{equation}
where dots stand for higher in $1/k$ corrections. Thus, the
renormalized
current forms the affine algebra at level $k$ (at least in the large
$k$
limit).

Finally we arrive at the conclusion that the classical stress-tensor
in eq.
(6.80) at the conformal point $S^{*}$ gets renormalized to the
operator
\begin{equation}
T(\tilde S^{*})={1\over k+c_V}:\tilde{\cal J}^2:,\end{equation}
with $\tilde{\cal J}$ the renormalized classical affine current.
In other words, the quantity $\hat J$ goes to $\tilde{\cal J}$ at the
conformal
point $S^{*}$. One can check that
\begin{equation}
{-\delta^{ab}\over k+2c_V}~+~4(k+2c_V)\tilde S^{a\bar a}\tilde
S^{b\bar
a}\stackrel{\tilde S\to\tilde
S^{*}}{\longrightarrow}{\delta^{ab}\over
k}~+~...\end{equation}
Therefore,
\begin{equation}
\hat J\stackrel{\tilde S\to\tilde S^{*}}{\longrightarrow}\tilde{\cal
J}.\end{equation}

Now let us return to the theory in eq. (5.49). In the latter, the
classical
affine current forms the algebra with level $k$. Hence, the given
classical
current is not needed to undergo any renormalization to emerge in the
Sugawara
construction. This is also true because the large $k$ limit is the
strong
coupling limit for the given theory. The renormalization affects only
the
coefficient matrix in the stress-tensor:
\begin{equation}
T(L)=L_{ab}:{\cal J}^a{\cal J}^b:, \end{equation}
where ${\cal J}$ is the affine current with level $k$, whereas
$L_{ab}$ is a
matrix related to the coupling constant matrix $S^{a\bar a}$. The
given
construction does not contradict to the renormalizability of the
classical
stress-tensor. Indeed, for the antiholomorphic component of the
conjectured
stress-tensor, one obtains
\begin{equation}
\bar T( L)= L\left[\bar{\tilde J}^2~+~4k\bar{\tilde J}\tilde
g_1(\mbox{Tr}\tilde S\bar{\tilde J}_{\tilde h_0})\tilde
g^{-1}_1~+~4k^2(\mbox{Tr}\tilde S\bar{\tilde J}_{\tilde
h_0})^2\right].\end{equation}
One can see that the expression in eq. (6.100) has the same terms as
the
classical $\bar T(S)$ only with different coefficients. Therefore,
the
renormalizability allows us to consider the operator in eq. (6.100)
as the
quantum renormalized component of the energy stress-tensor at
conformal points.
The condition of conformal invariance amounts to imposing the
Virasoro algebra
on $\bar T( L)$. This fixes values of the matrix $ L$ \cite{Halpern}.
However,
this condition does not tell us how the matrix $L$ is related to the
coupling
constant matrix $S$. This relation has to be derived from dynamics.
In the
large $k$ limit, one can find
\begin{equation}
L=-{I\over k}~+~{2c_V\over k}\hat S^2~+~...\end{equation}
The left hand side of eq. (6.101) can be found from the
master-Virasoro
equation which describes embeddings of the Virasoro algebra into the
affine
algebra \cite{Halpern}. As a result, the system of two interacting
WZNW models
may be used for Lagrangian interpretation of the algebraic
affine-Virasoro
construction.

All in all we have established that equation (6.101) (or its
non-perturbative
generalization) represents a sufficient condition of the conformal
invariance
of the system of two interacting WZNW models. We have checked in
\cite{Gates}
that at $k=4$ the system of two interacting $SL(2)$ WZNW models is
conformal
with a continuous coupling constant. Similarly, the affine-Virasoro
construction on $SL(2)$ also has a continuous solution at $k=4$
\cite{Morozov}.
This is a non-trivial evidence that relation (6.101) is also a
necessary
condition of the conformal symmetry. Some non-perturbative arguments
in favour
of it have been presented in \cite{Soloviev4} as well.

\section{Conclusion}

We have exhibited that the system of two interacting WZNW models has
the weak
coupling phase and the strong coupling phase. The scaling properties
of the
theory in these two phases are completely different. We have
presented both
perturbative and non-perturbative arguments that in the strong
coupling phase
the theory flows to non-trivial conformal points, which may
correspond to some
of the string compactifications. The hope is that the given
properties of
interacting WZNW models will turn out to be useful for understanding
the string
theory symmetries.

{\bf Acknowledgements:}

I am indebted to C. Hull and J. Gates for interesting discussions.
I would like to thank the British PPARC and the Physics Department of
the
University of Maryland for financial support.

\end{document}